\documentclass[aps,twocolumn,superscriptaddress,amsmath,nofootinbib,longbibliography,amssymb,prx,10pt]{revtex4-2}
\usepackage{graphicx}
\usepackage{float}
\usepackage[normalem]{ulem}
\usepackage{dcolumn}
\usepackage{bm}
\usepackage{physics}
\usepackage{wasysym}
\usepackage{comment}
\usepackage{placeins} 
\usepackage{xcolor, colortbl}
\usepackage[ruled]{algorithm2e}
\definecolor{darkblue}{rgb}{0,0,0.6}
\usepackage[colorlinks,linkcolor=darkblue,citecolor=darkblue,urlcolor=darkblue]{hyperref}
\newcommand{\Dr}{D_{\rm rot}}

\newcommand{\rev}[1]{{\color{black}#1}}

\begin{document}

\title{How to quantify long-time rotational motion in molecular systems}

\author{Romain Simon}

\affiliation{Laboratoire Charles Coulomb (L2C), Universit\'e de Montpellier, CNRS, 34095 Montpellier, France}

\author{Hadrien Bobas}

\affiliation{Gulliver, CNRS UMR 7083, ESPCI Paris, PSL Research University, 75005 Paris, France}

\author{Fran\c{c}ois Villemot}

\affiliation{Gulliver, CNRS UMR 7083, ESPCI Paris, PSL Research University, 75005 Paris, France}

\author{Jean-Louis Barrat}

\affiliation{Universit\'e Grenoble Alpes, CNRS, LIPhy, 38000 Grenoble, France}

\author{Ludovic Berthier}

\email{ludovic.berthier@espci.fr}

\affiliation{Gulliver, CNRS UMR 7083, ESPCI Paris, PSL Research University, 75005 Paris, France}

\date{\today}

\begin{abstract}
We show that all existing methods quantifying rotational motion in molecular fluids eventually have severe limitations in systems undergoing complex rotational motion characterised by slow, heterogeneous, or intermittent dynamics. This impacts in particular the study of rotational dynamics in molecular supercooled liquids near their glass transition, as well as discussions of the decoupling between rotational and translational motion and violations of the Debye-Stokes-Einstein relation. We present a brief overview of existing methods and explain why none of them can accurately capture the evolution of rotational dynamics from a diffusive fluid to an arrested solid, thus resolving inconsistent literature results. We then introduce an empirical method that efficiently solves all issues. We benchmark our method devising a family of continuous time random walk models for rotational dynamics. Our method correctly quantifies the statistics of free and caged rotational motion, as well as non-Gaussian and non-Fickian rotational dynamics, and should allow a better characterisation of dynamic heterogeneity in the rotational motion of supercooled molecular fluids. 
\end{abstract}

\maketitle

\section{Introduction}

\label{sec:introduction}

In statistical physics, the macroscopic physical behavior of a system is explained via the detailed analysis of the microscopic degrees of freedom. To understand the dynamic properties of molecular systems, such as transport properties, one should analyze the microscopic dynamics of the molecules~\cite{boon1980molecular,barrat2003basic}. Generally speaking, molecules are three-dimensional objects characterised by six degrees of freedom for its rigid body translations and rotations, and many more if the molecules contains internal degrees of freedom~\cite{debye1929polar,goldstein2002classical,boon1980molecular}. Here, we consider the three rotational degrees of freedom of rigid molecules of arbitrary shapes, that we treat as three-dimensional rigid bodies~\cite{favro1960theory}. In that case, the rotational dynamics can be seen as a random walk in angular space, obtained from the composition of infinitesimal rotational motions, that can be analytically described by rotation matrices. Our goal is to quantify the detailed statistics of rotational motion emerging at long times in molecular fluids, in particular when the system displays slow, heterogeneous and intermittent dynamics for which there can exist a broad range of timescales in which case the assumption of Brownian motion with Gaussian statistics breaks down~\cite{ediger2000spatially,Richert2002heterogeneous}.

Our central motivation to revisit such classic problem stems from the study of deeply supercooled molecular fluids approaching their glass transition~\cite{ediger2000spatially,Richert2002heterogeneous,Berthier2011}. In that case, the molecular dynamics slows down dramatically, and it is generally expected that an analytic description of these dynamics in terms of a free angular diffusion model becomes inadequate~\cite{Richert2002heterogeneous}. There are many experimental~\cite{cicerone1995molecules,ediger2000spatially,Richert2002heterogeneous,paeng2015ideal} and numerical~\cite{berthier2011dynamical} indications that molecular motion becomes intermittent, characterised by broad distribution of timescales, leading to possible decoupling between rotational and translational motion~\cite{fujara1992translational,stillinger1994translation} and violations of Debye-Stokes-Einstein relation between transport coefficients~\cite{tarjus1995breakdown,cicerone1996enhanced,andreozzi1997evidence,ediger2000spatially,mandel2022single}. Superficially, these questions appear well-documented, but we have recently discovered,  during the course of a molecular dynamics study~\cite{simon2026molecular}, that available methods to quantify rotational motion all suffer from severe limitations and may even lead to physically inconsistent results. These observations have motivated the present theoretical analysis. 

In general, rotational motion can be analysed via two qualitatively different approaches. 

In the first approach, the information about rotational motion is inferred by observing the molecule at two discrete  times~\cite{berne1976dynamic,hansen1986theory}. The molecule is first observed at an arbitrary initial time, $t=0$, and at a consecutive time $t>0$ later. If these observations are done in a numerical simulation, one may be able to reconstruct a $3 \times 3$ rotation matrix, $R(t)$, that would transport the molecule from its initial position at $t=0$ to the final one at $t$. Equivalently, this defines an angular displacement that can be represented by a three-dimensional rotation vector, $\boldsymbol{\phi}(t)$ (see Sec.~\ref{sec:methods} for precise definitions). While these two objects are relatively easily extracted in a simulation, experimental techniques~\cite{berne1976dynamic,lunkenheimer2000glassy,flamig2020nmr,Torre1998time,hinze2000orientational} such as dielectric spectroscopy, depolarized dynamic light scattering, nuclear magnetic resonance \rev{or optical Kerr effect experiments,} determine angular correlation functions that contain an equivalent information about molecular rotations evaluated at two times $t=0$ and $t$~\cite{hansen1986theory}. \rev{In dielectric experiments for instance, one has access, when cross correlations are negligible~\cite{henot2023orientational}, to the time correlation $C_1(t) = \langle P_1( \cos \theta(t) ) \rangle$ where $\theta(t)$ represents the amplitude of the rotation of the dipolar moment and $P_1(x)$ the Legendre polynomial of rank 1. In this example, $\theta(t)$ is directly related to the projection of the rotation vector $\boldsymbol{\phi}(t)$ along the direction of the applied electric field.}

In this first family of methods, the angular displacement is necessarily bounded because $| \boldsymbol{\phi}(t) | \in [0, \pi]$, by construction. This makes sense, since the position of the molecule at time $t$ is the same whether it made one or several turns in the time interval $[0,t]$. \rev{In addition, functions such as $C_1(t)$ are unaffected if $\theta(t) \to \theta(t) + 2 \pi$.} As a result, the total angular displacement is also necessarily bounded, $\lim_{t \to \infty} \langle | \boldsymbol{\phi}(t) - \boldsymbol{\phi}(0)|^2 \rangle \to const$. There exists no long-time Fickian limit from which a rotational diffusion constant, $\Dr$, can be extracted. \rev{In dielectric experiments, it is therefore customary to define instead a timescale $\tau_1$ from the time decay of $C_1(t)$ to quantify the rotational dynamics. It is then sometimes assumed that $\Dr \propto \tau_1^{-1}$, with the implicit hypothesis that the motion of the dipolar moment can be described by a simple Brownian motion on the unit sphere~\cite{favro1960theory}. A central issue arises with these approaches when the rotational dynamics only becomes Gaussian and Fickian at long times, and the underlying assumptions break down.}  

A second family of approaches can be developed when the molecules can be tracked at all times over the interval $[0,t]$, as is possible in computer simulations or microscopy experiments of colloidal systems. In that case, the idea is to define an angular displacement between times $0$ and $t$ that accumulates infinitesimal angular displacements along the entire trajectory~\cite{ka_dynamics_1997,wong2008evaluating,rozmanov2012transport,zhang2023dynamic,zendehroud2024molecular,holtbrugge2025robust}. This seems to make intuitive sense, as this opens the possibility to distinguish between molecules that make one or several turns. By analogy with translational motion where the total displacement is the time integral of the instantaneous translational velocity, a total angular displacement $\boldsymbol{\phi}(t)$ defined as the time integral of the angular velocity was analysed in many previous works, in particular in the context of glass transitions studies~\cite{ka_dynamics_1997,leporini2001viscous,bagchi_anomalous_2005,mazza2006relation,lombardo2006computational,becker2006fractional,chong2009coupling,edmond_decoupling_2012,sese_rotational_2012,kawasaki_spurious_2019,agosta2021supercooled,jin2024understanding,ren2025stokes}. By construction, this total rotational displacement is no longer bounded and can potentially keep track of multiple $2\pi$ rotations (think of the simple case of a spinning object). As a result, this integral method can be used to define a mean-squared angular displacement that becomes Fickian in the long-time limit, 
\begin{equation}
  \lim_{t \to \infty} | \boldsymbol{\phi}(t) - \boldsymbol{\phi}(0)|^2 = 3 \Dr t,
\label{eq:fickian}
\end{equation}
that defines $\Dr$ as a long-time rotational diffusion constant. This approach has been used extensively in the literature on supercooled liquids, even though some have noted that this observable may lead to results that are difficult to interpret~\cite{kawasaki_spurious_2019}, or that are interpreted as challenging the very concept of a rotational diffusion constant~\cite{chong2009coupling}. We will show that the integral method is in fact mathematically inconsistent, because it incorrectly assumes that three-dimensional rotation matrices commute. As a result, the time integral accumulates small errors that eventually prevent the correct evaluation of $\Dr$ at large times. 

In this paper, we demonstrate that in molecular systems where rotational dynamics is slow or intermittent, the two \rev{families} of existing approaches have such severe limitations that they fail to capture the correct value and temperature evolution of a long-time rotational diffusion constant. This problem is particularly acute for supercooled molecular liquids, where both types of measurements completely fail, and thus they provide an incorrect physical picture of the relation between rotational and translational motion at the molecular scale and of the evolution of rotational dynamics across the glass transition. \rev{The method and analysis we propose here are specifically aimed at computer simulations, where the orientation of molecules is accurately known at all times. Our goal is to extract physically correct and accurate physical observables to quantify rotational dynamics at large times.}  

To support these claims, we introduce a family of relatively simple models for angular rotational motion that are inspired by continuous time random walk models that were initially created to describe translational diffusion in complex media~\cite{montroll1965random}. By carefully creating these analytic models, we will expose the limits and problems encountered with conventional methods, and show their inability to follow the gradual transition between a diffusive fluid to an arrested solid. We will then propose an empirical method that allows us to extract a rotational diffusion constant for systems displaying complex orientational dynamics. We demonstrate its applicability in a series of models of increasing complexity that mimic the slow and heterogeneous dynamics of supercooled molecular fluids in both liquid and glass phases. We conclude that the concept of a long-time rotational diffusion constant can be valid in supercooled molecular fluids, but that it was not properly accessed in earlier numerical simulations. In addition, it cannot be easily determined using scattering experiments but could be determined in confocal microscopy studies of colloidal systems. We discuss the implications of our work, that should be broadly applicable to revisit the molecular dynamics of fluids characterised by slow and anomalous rotational dynamics. 

In Sec. \ref{sec:methods} we introduce the formalism and available methods to keep track of rotational motion of three-dimensional objects. In Sec.~\ref{sec:conventional}, we demonstrate the limits of both families of conventional methods. In Sec.~\ref{sec:novel}, we introduce a new way to study rotational motion. In Sec.~\ref{sec:application} we validate this method using a series of simple models for rotational motion. We conclude and discuss perspectives for future work in Sec.~\ref{sec:conclusion}. 

\section{Kinematics of rotational dynamics}

\label{sec:methods}

\subsection{Rotation matrices and rotation vectors}

Let us consider a rigid body in three dimensions ($d=3$). We attach to it a body-fixed Cartesian coordinate system, denoted $XYZ(t)$, which evolves with time, and compare it to a space-fixed Cartesian system, $xyz$. A rigid body has six degrees of freedom, three for translations and three for rotations. We focus exclusively on the rotational degrees of freedom and disregard translations. For convenience, we choose the space-fixed  axes $xyz$ to coincide with the body-fixed axes $XYZ(0)$ at $t=0$, so that both frames are initially aligned. By convention, the axes of $xyz$ coincide with the standard basis vectors $\mathbf{e}_x = {(1,0,0)}^\top$, $\mathbf{e}_y = {(0,1,0)}^\top$, and $\mathbf{e}_z = {(0,0,1)}^\top$.

The orientation of the rigid body at time $t$ can be represented by a rotation matrix, $R$ that belongs to the special orthogonal group $\mathrm{SO}(3)$. The rotation matrix maps a given vector expressed in the body-fixed frame $XYZ(t)$ to its representation in the space-fixed frame $xyz$. Equivalently, the columns of the matrix $R$ are the coordinates of the body-fixed axes expressed in the space-fixed basis. Since $R \in \mathrm{SO}(3)$, it satisfies~\cite{hall_lie_2015}
\begin{equation}
    RR^\top = I, \qquad \det(R) = 1.
\end{equation}

An alternative, and often more intuitive, representation of three-dimensional rotations is the axis-angle formulation~\cite{forstner2016photogrammetric}, which follows directly from Euler's rotation theorem~\cite{euler1953commentationes}. Any rotation can also be described by a rotation angle $\theta \in [0,\pi]$ about a unit vector $\mathbf{n} \in \mathbb{R}^3$, with $| \mathbf{n} | = 1$. The corresponding rotation vector (or Euler vector) is defined as
\begin{equation}\label{eq:defeulervector}
    \boldsymbol{\Omega} = \theta \mathbf{n}.
\end{equation}
Other formalisms to represent rotations exist, such as Euler angles~\cite{weisstein2009euler} and quaternions~\cite{mukundan2012quaternions}, but they are not required for the present discussion, as they contain equivalent information that would lead to similar conclusions. 

\begin{figure}
\includegraphics[width=\columnwidth]{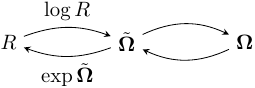}
    \caption{Schematic representation of the conversion from a rotation matrix $R$ to the skew-symmetric matrix representation of the corresponding rotation vector $\tilde{\boldsymbol{\Omega}}$ via the logarithm mapping and the inverse conversion via the exponential mapping. The rotation vector $\boldsymbol{\Omega}$ is easily extracted from its skew-symmetric representation $\tilde{\boldsymbol{\Omega}}$ by reading the matrix entries, and vice-versa.}
\label{fig:conversions}
\end{figure}

To connect the vector and matrix representations, we introduce the skew-symmetric operator
\begin{equation}
    \tilde{(\cdot)}: \mathbb{R}^3 \to \mathfrak{so}(3),
\end{equation}
where $\mathfrak{so}(3)$ denotes the Lie algebra of $\mathrm{SO}(3)$, consisting of $3 \times 3$ skew-symmetric matrices~\cite{hall_lie_2015}. For a given Euler vector $\boldsymbol{\Omega}={(\Omega_x,\Omega_y,\Omega_z)}^\top$, we define the operator as
\begin{equation}
  \tilde{\boldsymbol{\Omega}} =
    \begin{bmatrix} 
        0 & -\Omega_z & \Omega_y \\ 
        \Omega_z & 0 & -\Omega_x \\ 
        -\Omega_y & \Omega_x & 0 
    \end{bmatrix}.
\end{equation}
This mapping is bijective, meaning that for any skew-symmetric matrix $\tilde{\boldsymbol{\Omega}} = {[ \tilde{\Omega}_{ij} ]} \in \mathfrak{so}(3)$,  there exists a corresponding rotation vector 
$\boldsymbol{\Omega}$:
\begin{equation}
    \boldsymbol{\Omega} = 
    ( -\tilde{\Omega}_{23}, \tilde{\Omega}_{13}, -\tilde{\Omega}_{12} )^\top.
\label{eq:recovervector}
\end{equation}
We summarize in Fig.~\ref{fig:conversions} the conversion methods from matrix to vector rotations, and the inverse path.

\subsection{Converting a rotation vector to a rotation matrix}

The relationship between a rotation vector $\boldsymbol{\Omega}=\theta\mathbf{n}$ and its corresponding rotation matrix $R$ is given by the matrix exponential~\cite{hall_lie_2015}:
\begin{equation}
    R = \exp  \tilde{\boldsymbol{\Omega}} .
    \label{eq:exp}
\end{equation}
This exponential map provides the fundamental link between $\mathfrak{so}(3)$ and $\mathrm{SO}(3)$. Although $\exp  \tilde{\boldsymbol{\Omega}} $ can be computed via its Taylor series expansion, a closed-form and numerically-stable alternative is given by the Rodrigues rotation formula~\cite{rodrigues1840lois}:
\begin{equation}
    R = I_3 + \sin\theta \, \tilde{\mathbf{n}} + (1 - \cos\theta)\, \tilde{\mathbf{n}}^2,
    \label{eq:Rodri}
\end{equation}
where $\tilde{\mathbf{n}} = \tilde{\boldsymbol{\Omega}}/\theta$ is the skew-symmetric matrix associated with $\mathbf{n}$.

\subsection{Converting a rotation matrix to a rotation vector}

\label{sec:matrixtovector}

Conversely, given a rotation matrix $R \in \mathrm{SO}(3)$, the corresponding rotation vector $\boldsymbol{\Omega}$ can be extracted from the skew-symmetric matrix $\tilde{\boldsymbol{\Omega}}$, that is itself obtained using the matrix logarithm:
\begin{equation}
    \tilde{\boldsymbol{\Omega}} = \log R.
\end{equation}
As with the exponential, computing the logarithm directly is possible but numerically sub-optimal. An efficient alternative is described  in Ref.~\cite{liang2018efficient} and proceeds as follows. First, the axis of rotation $\mathbf{n}$ is obtained as the eigenvector of $R$ with eigenvalue $1$:
\begin{equation}
    R\mathbf{n} = \mathbf{n}.
\end{equation}
Then, the cosine and sine of the rotation angle $\theta$ are computed from
\begin{align}
    \cos \theta &= \frac{1}{2}\,\bigl(\mathrm{Tr}(R) - 1\bigr), \\
    \sin \theta &= -\frac{1}{2}\,\mathrm{Tr}(\tilde{\mathbf{n}}R).
\end{align}
The angle $\theta$ is then recovered using the two-argument arctangent function $\mathrm{atan2}(y, x)$, that returns the argument of the complex number $x+iy$:
\begin{equation}
    \theta = \mathrm{atan2} (\sin \theta, \cos \theta ).
\end{equation}
Finally, the rotation vector is reconstructed, $\boldsymbol{\Omega} = \theta \mathbf{n}$. Although this procedure differs from the conventional $\arccos$-based extraction method,  it provides improved numerical stability, particularly when $\cos \theta$ is close to $\pm 1$.

\subsection{Combining successive rotations}

\label{sec:combining}

Consider two successive rotations of a rigid body, represented by two rotation vectors  $\boldsymbol{\Omega}_1$ and $\boldsymbol{\Omega}_2$, both expressed in the two consecutive body-fixed frames $XYZ$. The overall rotation matrix corresponding to the composition of these two rotations is given by
\begin{equation}
\label{eq:rotmatmult}
    R = \exp \tilde{\boldsymbol{\Omega}}_1  \cdot \exp \tilde{\boldsymbol{\Omega}}_2 ,
\end{equation}
where $\tilde{\boldsymbol{\Omega}}_i \in \mathfrak{so}(3)$ denotes the skew-symmetric matrix associated with the rotation vector $\boldsymbol{\Omega}_i$. As the overall rotation matrix lies in $\mathrm{SO}(3)$, it can be expressed as $R = \exp \tilde{\boldsymbol{\Omega}}$, with $\boldsymbol{\Omega}$ the total rotation vector.

The order of the multiplication in Eq.~(\ref{eq:rotmatmult}) is crucial. Since $\mathrm{SO}(3)$ is a non-Abelian Lie group, the corresponding rotation matrices do not commute in general, so that
\begin{equation}
    \exp \tilde{\boldsymbol{\Omega}}_1  \cdot \exp \tilde{\boldsymbol{\Omega}}_2  
    \neq \exp \tilde{\boldsymbol{\Omega}}_2 \cdot \exp \tilde{\boldsymbol{\Omega}}_1 .
\end{equation}
This directly implies that $R$ is not equal to $\exp ( \tilde{\boldsymbol{\Omega}}_1 + \tilde{\boldsymbol{\Omega}}_2)$. The total rotation vector is therefore not given by the sum the individual rotation vectors,
\begin{equation}
  \boldsymbol{\Omega} \neq \boldsymbol{\Omega}_1+\boldsymbol{\Omega}_2 .
\end{equation}

By induction, for $N$ successive rotations of the rigid body, defined by rotation vectors $\boldsymbol{\Omega}_1, \dots, \boldsymbol{\Omega}_N$ in the body-fixed frame, the total rotation matrix is
\begin{equation}
  R = \exp \tilde{\boldsymbol{\Omega}}_1 \cdot \exp \tilde{\boldsymbol{\Omega}}_2 \cdot \, \cdots \, \cdot \exp \tilde{\boldsymbol{\Omega}}_N,
  \label{eq:sum}
\end{equation}
with each $\tilde{\boldsymbol{\Omega}}_i \in \mathfrak{so}(3)$ corresponding to the skew-symmetric representation of each $\boldsymbol{\Omega}_i$. By the same reasoning the total rotation vector $R$ is not equal to the sum of the individual rotation vectors,
\begin{equation}
  \label{eq:rotisnotsum}
    \boldsymbol{\Omega} \neq \sum_{i=1}^{N}\boldsymbol{\Omega}_i , 
  \end{equation}
but is instead obtained by converting $R$ from Eq.~(\ref{eq:sum}) into its corresponding Euler vector using the formalism in Sec.~\ref{sec:matrixtovector}.

\subsection{Two simple numerical models: free and confined angular random walks}

\label{sec:tests}

To test the validity and accuracy of various definitions of a total angular displacement, we first devise two simple numerical models. In essence, these models generate a dynamic trajectory for a body-fixed coordinate system $XYZ(t)$, by combining small successive rotations and starting from the initial condition $R(t=0) = I_3$.

\begin{figure}
\includegraphics[width=\linewidth]{figures/fig2.png}
\caption{Simple models for (a) a free and (b) a confined angular random walk, as introduced in Sec.~\ref{sec:tests}. The parameters are $\tau=1$ for the distribution of jump times, $\epsilon=0.05$ for their maximal amplitude, and $\theta_c=0.2$ for the confining angle in (b). The trajectory in (c) is the combination of cage dynamics and intermittent jumps, as introduced in Sec.~\ref{sec:intermittent}. The cage parameters are identical to (b) and the jump parameters are $\epsilon_j = 0.25 = 5 \epsilon$, and $\tau_j=400$. In all panels the color codes for the time.} 
\label{fig:sketch}
\end{figure}

In the first model, we generate a free angular random walk where a small rotation vector $\delta \boldsymbol{\phi}$ is applied in the rotating body-fixed frame, causing the rotation matrix at time $t + \delta t$ to evolve according to:
\begin{equation}
  \label{eq:rottime}
    R(t+ \delta t) = R(t) \cdot \exp  \delta \tilde{\boldsymbol{\phi}}(t) ,
\end{equation}
where $\delta \tilde{\boldsymbol{\phi}}(t)$ is the skew-symmetric matrix associated with the rotation vector $\delta \boldsymbol{\phi}(t)$. In the example shown in Fig.~\ref{fig:sketch}(a), we consider a continuous time random walk (CTRW) where the time intervals $\delta t$ are drawn from a Poissonian distribution, $P(\delta t) = \tau^{-1} \exp(-\delta t/\tau)$ with $\tau=1$, and the three components of the rotations vectors $\delta \boldsymbol{\phi}$ are independent random numbers drawn from the flat distribution $[- \epsilon, \epsilon]$, with a small amplitude $\epsilon=0.05$. This first model mimics the diffusive rotational dynamics of a free molecule. In the limit $\epsilon\rightarrow 0$, it corresponds to a Brownian motion in $\mathrm{SO}(3)$ that is discussed further in Appendix~\ref{app:FP}. This is physically analogous to the more usual rotational diffusion on the unit sphere~\cite{mazo2008}, but not equivalent, as the latter process only involves two independent degrees of freedom. A representative trajectory obtained for the free angular random walk is shown in Fig.~\ref{fig:sketch}(a).  

We devise a second model to mimic the physics expected for low temperature glassy systems where rotational motion may get frozen, the molecular orientation is now confined within a restricted cage, and the molecules never fully rotate. To this end, we generate trajectories using again Eq.~(\ref{eq:rottime}), but we add a constraint: if the angle $\theta(t+dt)$ corresponding to the total rotation matrix $R(t+dt)$ becomes larger than a threshold $\theta_c$, the small angular jump is rejected and $R(t+dt) \leftarrow R(t)$. In the example shown in Fig.~\ref{fig:sketch}(b) we have chosen an orientational constraint $\theta_c = 0.2$ using again $\tau=1$ and $\epsilon=0.05$. Notice that the cage constructed by this simple model does not correspond to a Gaussian distribution of angles around the average position, as would be generated by thermal fluctuations in an arrested solid. A simplified version of this model (diffusion of a rod restricted to a cone) was studied analytically before~\cite{wang1980time}.  

The combination of caged dynamics and intermittent jump events shown in Fig.~\ref{fig:sketch}(c) is introduced below in Sec.~\ref{sec:intermittent}. 

\section{Conventional methods and their limitations}

\label{sec:conventional}

The conventional methods can be organized into two broad categories. In the first one, the system is observed at times $t=0$ and $t$ and the rotational displacement between these two times is estimated. In the second family, the system is observed at all times between $t=0$ and $t$ and the rotational displacement is accumulated from infinitesimal displacements along the entire trajectory. In this section, we demonstrate that each approach may fail even in very simple cases, and this considerably impacts the measurement of a long-time rotational diffusion constant in systems with slow dynamics. 

\subsection{Using the total rotation vector between two times}

\label{sec:eulervector}

When the molecule is observed at times $t=0$ and $t>0$, the angular displacement $\boldsymbol{\phi}(t)$ between these two times can be estimated via the total rotation vector $\boldsymbol{\Omega}(t)$, also known as the Euler vector, as defined in Eq.~\eqref{eq:defeulervector}.

Writing $\boldsymbol{\Omega} = \theta \mathbf{n}$, with $\mathbf{n}$ a unit vector, it follows that $| \boldsymbol{\phi}(t) -\boldsymbol{\phi}(0) |^2  = |\boldsymbol{\Omega}(t)|^2 = \theta^2$. Since the rotation angle satisfies $\theta \in [0, \pi]$, this implies that $0 \leq \theta^2 \leq \pi^2$, and the total angular displacement is therefore bounded.

This is physically reasonable: when observing the molecule at times $0$ and $t$ only, it makes no difference whether the molecule performed one or multiple  rotations before arriving at its final position at time $t$. Consequently, this observable does not exhibit unbounded diffusive growth. In that case, a rotational diffusion constant can only be defined if the motion is Fickian much before the angular displacement reaches its upper limit. This obviously becomes problematic in situations where the rotational motion only becomes Fickian at very large times. 

\begin{figure}
\includegraphics[width=\columnwidth]{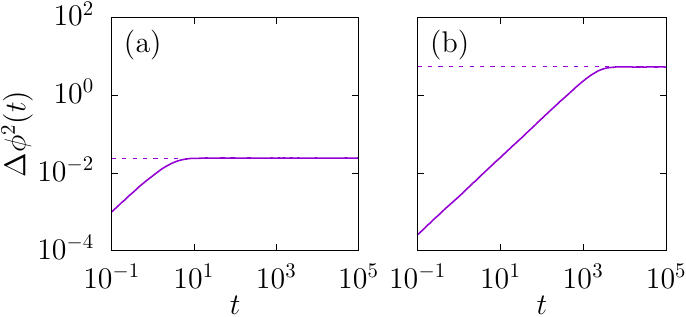}
\caption{Test of Euler vector method on (a) a confined and (b) a free angular random walk. In (b) no Fickian regime is reached at large times because the angular displacement is bounded, preventing the definition of $D_{
\rm rot}$ using Eq.~(\ref{eq:fickian}) for the diffusive model.}
\label{fig:eulervector}
\end{figure}

To illustrate this point, we show in Fig.~\ref{fig:eulervector} the mean-squared angular displacement
\begin{equation}
  \Delta \phi^2 (t) = \langle |\boldsymbol{\phi}(t) - \boldsymbol{\phi}(0) |^2 \rangle,
  \label{eq:MSD}
\end{equation}
where the brackets indicate an average over an ensemble of trajectories. We have measured this quantity for both the confined and the free angular random walks models introduced in Sec.~\ref{sec:tests} using $10^4$ trajectories. In both models, we use a Poissonian distribution of jump times with $\tau=1$, and reject jumps when $\theta(t) > \theta_c = 0.2$ for the confined model whereas all jumps are accepted for the free model. The results are displayed in Fig.~\ref{fig:eulervector}.  

For the confined model in Fig.~\ref{fig:eulervector}(a) we observe that $\Delta \phi^2 (t)$ grows diffusively at very short times and reaches a plateau a long times, whose value is set by $\theta_c$, as expected. However, the same qualitative behavior is observed for the free random walk model, see Fig.~\ref{fig:eulervector}(b). In that case, the long-time plateau does not result from the physics of the model but stems from the definition of the observable itself, which is bounded at long times. These results confirm the absence of a long-time Fickian limit for $\Delta \phi^2 (t)$ in all cases when using this approach. 

The two plateau values can be estimated analytically as they correspond to the uniform sampling of the rotation axis $\mathbf{n}$ on the unit sphere and a probability density function for $\theta$, denoted $p(\theta)$, of the form given in Ref.~\cite{miles1965random} (see Appendix~\ref{app:FP} for an alternative derivation):
\begin{equation}
  p(\theta) = \frac{2}{\theta_c-\sin \theta_c} \sin^2  \left( \frac{\theta}{2} \right),
  \label{eq:thetarandom}
\end{equation}
leading to the second moment
\begin{equation}
\langle \theta^2 \rangle = \frac{ \theta_c^3/3 +  (2-\theta_c^2) \sin \theta_c - 2 \theta_c \cos \theta_c }{\theta_c - \sin \theta_c} . 
\label{secondmoment}
\end{equation}
For small $\theta_c$, we get the expected scaling behavior $\theta^2 \approx \frac{3}{5} \theta^2_c $. The horizontal dashed lines in Fig.~\ref{fig:eulervector} confirm that $\Delta \phi^2 (t \to \infty) = \langle \theta^2 \rangle$, using respectively $\theta_c=0.1$ and $\theta_c=\pi$ in Eq.~(\ref{secondmoment}). 

For the free random walk with the chosen parameters, we expect the rotational diffusion constant to be given by $\Dr = \epsilon^2 / (3 \tau)$. While the relation $\Delta \phi^2(t) = 3 \Dr t$ is obeyed at short times in Fig.~\ref{fig:eulervector}(b), the long-time limit of $\Delta \phi^2(t)$ cannot be used to estimate the rotational diffusion coefficient. In addition, there is no timescale over which the distribution of angular displacement is a Gaussian, because it is a bounded variable. This demonstrates our central point: for more complex systems where a Fickian limit is only reached at large times and large displacements, $\Delta \phi^2 (t)$ may reach its asymptotic plateau ($\pi^2/3+2$, given by Eq.~(\ref{secondmoment})) long before the Fickian regime is entered, which can therefore never be studied using this approach.     

\subsection{Integration of the angular velocity along trajectory}

\label{sec:angularvelocity}

When the system can be observed at all times between $t=0$ and $t$, it is possible to define a total angular displacement as the sum of small increments~\cite{ka_dynamics_1997}. For a continuous time process this amounts to defining the angular displacement vector, $\boldsymbol{\phi}(t)$, as the time integral of the instantaneous angular velocity vector, $\boldsymbol{\omega}(t)$:
\begin{equation}
  \boldsymbol{\phi}(t) = \int_0^t \boldsymbol{\omega}(t') \, dt' .
  \label{eq:velocityintegral}
\end{equation}
In the present context, the angular velocity vector, $\boldsymbol{\omega}(t)$, is defined by the infinitesimal rotation vector, $d \boldsymbol{\phi}(t)$, which maps the orientation of a body-fixed frame, $XYZ(t)$, to its subsequent orientation at $t+dt$, namely $\boldsymbol{\omega}(t) = d \boldsymbol{\phi}(t)/dt$, and Eq.~\eqref{eq:velocityintegral} can be alternatively expressed as:
\begin{equation}
  \boldsymbol{\phi}(t) =  \int_0^t d \boldsymbol{\phi}(t').
  \label{eq:rotationintegral}
\end{equation}
This formulation provides a direct method for tracking the cumulative rotational motion of a molecule over time. The use of the integrated angular velocity as an observable for molecular rotation has been a subject of extensive investigations, particularly in the study of supercooled liquids and glass-forming systems~\cite{leporini2001viscous,chong2009coupling, edmond_decoupling_2012, ka_dynamics_1997, kawasaki_spurious_2019, lombardo2006computational, sese_rotational_2012, mazza2006relation, bagchi_anomalous_2005, becker2006fractional, edmond_decoupling_2012,agosta2021supercooled,jin2024understanding,ren2025stokes}. Finally, for a discrete time process composed of small increments, Eq.~(\ref{eq:rotationintegral}) becomes a discrete sum
\begin{equation}
  \boldsymbol{\phi}(t) = \sum_i \delta \boldsymbol{\phi}(t_i).
  \label{eq:timesum}
\end{equation}

\begin{figure}
  \includegraphics[width=\linewidth]{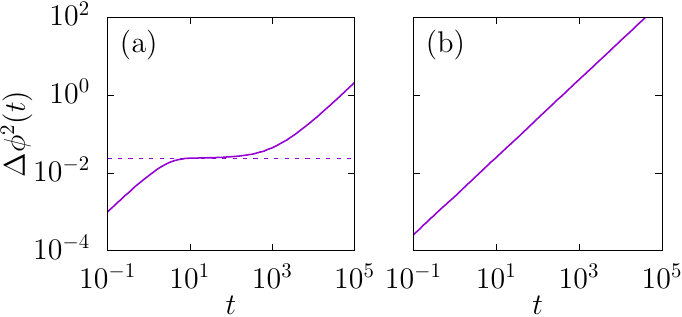}
  \caption{Test of integral method on (a) a confined and (b) a free angular random walk. In (a) an erroneous Fickian regime appears due to an accumulation of small errors.}
\label{fig:testofintegral}
\end{figure}

For the two models discussed in Sec.~\ref{sec:tests}, we use Eq.~(\ref{eq:timesum}) to obtain $\boldsymbol{\phi}(t)$ for each trajectory, and the average over trajectories then yields $\Delta \phi^2(t)$ via Eq.~(\ref{eq:MSD}). The results are shown in Fig.~\ref{fig:testofintegral}. They are qualitatively very different from those obtained with the Euler vector in Fig.~\ref{fig:eulervector}, for both our simple models.

For the confined angular random walk, the data in Fig.~\ref{fig:testofintegral}(a) reflect the correct physics at short times, with a diffusive initial growth, followed by a plateau regime, with a height that agrees with the expected analytic prediction. However, the long-time dynamics is diffusive again. This is surprising at first sight, as the trajectories are all confined in a small angular sector, as shown in Fig.~\ref{fig:sketch}(b). This result demonstrates that, for this integral method, an apparent Fickian regime is obtained. This would then define a finite rotational diffusion coefficient, $\Dr>0$, although we expect that $\Dr=0$ for the confined model.

Turning to the free angular random walk in Fig.~\ref{fig:testofintegral}(b), the mean-squared angular displacement is Fickian over all timescales (expect perhaps at very short times due to the small number of steps) and this defines a finite rotational diffusion coefficient $\Dr>0$, which agrees with its theoretical prediction $\Dr = \epsilon^2 / (3 \tau)$.

These results show that the angular displacement defined by an integral approach  always leads to a Fickian behavior at large times with a rotational diffusion coefficient that can be estimated correctly or incorrectly, depending on the model. This is of course inconvenient. 

Mathematically, this failure is in fact very natural. In the integral approach, the total angular displacement is estimated via Eq.~(\ref{eq:timesum}), whereas in the original model the total rotation matrix is given by
\begin{equation}
  R (t) = \prod_i \exp  d \tilde{\boldsymbol{\phi}}(t_i) ,
\label{eq:Rsum}
\end{equation}
with $d \tilde{\boldsymbol{\phi}}(t_i)$ the skew-symmetric matrix associated to the rotation vector $d \boldsymbol{\phi} (t_i)$. As anticipated in Sec.~\ref{sec:combining}, the rotation matrix in Eq.~(\ref{eq:Rsum}) is not equal to $\exp  \tilde{\boldsymbol{\phi}}(t) $. In other words, the total angular displacement $\boldsymbol{\phi}$ estimated in Eq.~(\ref{eq:timesum}) is unrelated to the correct rotation matrix $R(t)$. 

In practice, a small error is made in each increment of the sum in Eq.~(\ref{eq:timesum}), which appears as an incorrect simplification of the true motion that is instead described by Eq.~(\ref{eq:Rsum}). This error can be estimated via the Baker-Campbell-Hausdorff formula~\cite{hausdorff1906symbolische}. For the combination of two small displacements of typical amplitude $\epsilon$, one has
\begin{equation}
  \tilde{\boldsymbol{\phi}}(t_1+t_2) = d \tilde{\boldsymbol{\phi}}(t_1) + d \tilde{\boldsymbol{\phi}}(t_2)  +  \frac{1}{2} [ d \tilde{\boldsymbol{\phi}}(t_1), d \tilde{\boldsymbol{\phi}}(t_2) ] + {\cal O} (\epsilon^3),
  \label{eq:BCH}
\end{equation}
where the square brackets denote the matrix commutator. This formula makes it clear that the estimate of the angular displacement as an integral would be correct if rotation matrices were commuting objects. In general, this is however not the case. We checked that including higher-order terms in Eq.~(\ref{eq:BCH}) only reduces the error, but this does not solve the problem. 

The accumulation of small errors at each step has dramatic consequences for the case of a confined angular random walk. Whereas the true rotation matrix remains bounded by the cage angle $\theta_c$, the total angular displacement evaluated as an integral is dominated at large times by the accumulation of small errors that results in an apparent Fickian behavior in $\Delta \phi^2(t) \propto t$, which is a physically incorrect result in that case. More broadly, we conclude that the integral method is unable to capture the evolution of the rotational diffusion constant across a fluid to solid transition. See more data on this point in Sec.~\ref{sec:intermittent} below.  

\section{A novel method to quantify long-time rotational motion}

\label{sec:novel}

We have examined two distinct definitions of a total angular displacement that have advantages and limitations. The total Euler vector correctly captures the confined angular random walk but never approaches a Fickian limit at large times for the diffusive case. On the other hand the integral approach correctly captures a Fickian limit for the free angular random walk, but is dominated by an accumulation of errors for the confined model.   

We would like to introduce a definition of a total angular displacement that would correctly capture the physics of both confined and free angular random walks. If successful, we could then attack problems where Fickian dynamics only arises at large times, for which unbounded angular displacements are needed, or systems undergoing a transition from free diffusion to confined motion. 

\begin{figure}
  \includegraphics[width=\columnwidth]{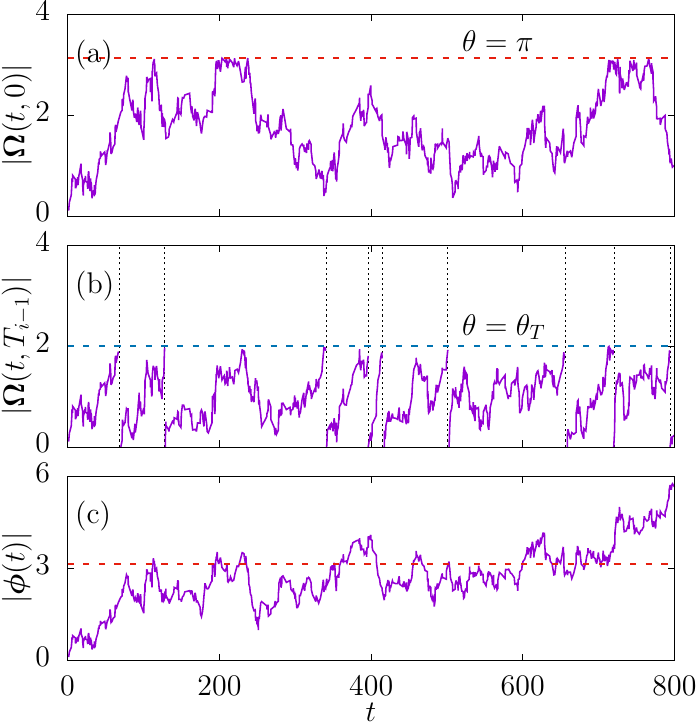}
  \caption{(a) The time evolution of the amplitude of the total Euler vector $|\boldsymbol{\Omega}(t,0)|$ representing the rotation matrix between times $0$ and $t$ leads to a bounded angular displacement $0\leq \theta(t) \leq \pi$.
  (b) When an angular threshold $\theta_T=1$ is introduced (horizontal dashed line), the amplitude of the Euler vectors $|\boldsymbol{\Omega}(T_{i-1},T_i)|$ is now computed in the time intervals between threshold crossings (vertical dashed lines). 
(c) The accumulated angular displacement $| \boldsymbol{\phi} (t)|$ defined in Eq.~(\ref{eq:new}) leads to an unbounded angular displacement that no longer saturates to $\pi$ (dashed line).}
  \label{fig:method}
\end{figure}

To this end, we introduce a new parameter, a threshold angular displacement, $\theta_T$, that will be adjusted empirically, and is used as follows. See Fig.~\ref{fig:method} for a concrete illustration in the case of a free angular random walk using the parameters $\tau=1$ and $\epsilon=0.25$. Starting from the usual initial conditions $R(t=0) = I_3$, we initially construct an angular displacement using the Euler vector $\boldsymbol{\phi}(t) = \boldsymbol{\Omega}(t,0)$ associated with the rotation matrix $R(t,0)$ between times $0$ and $t$. This definition is used up until a time $T_1$ defined as the first time when the angular displacement becomes larger than the threshold, $|\boldsymbol{\Omega} (t=T_1,0)| = \theta_T$. Up to $T_1$ our method is thus identical to the one using the total Euler vector described in Sec.~\ref{sec:eulervector}. 
 
The difference arises for $t>T_1$. For times $t>T_1$, the angular displacement is now estimated by first constructing the rotation matrix $R(t,T_1)$ between times $T_1$ and $t$, and its associated Euler vector $\boldsymbol{\Omega}(t,T_1)$. We then define the total angular displacement vector as $\boldsymbol{\phi}(t) = \boldsymbol{\Omega}(t,T_1) + \boldsymbol{\Omega}(T_1,0)$. This new definition is used until a time $T_2$ when the threshold is again crossed, $|\boldsymbol{\Omega} (t=T_2,T_1)| = \theta_T$. For $t>T_2$ the rotation matrix $R(t,T_2)$ is considered and the process is continued until the passage times $T_3, T_4, \cdots$ are crossed. After $n$ such crossings, the total angular displacement vector is given by
\begin{equation}
  \label{eq:new}
  \boldsymbol{\phi}(t) = \boldsymbol{\Omega} (t, T_n) + \sum_{i=1}^n \boldsymbol{\Omega}(T_{i}, T_{i-1})  ,
\end{equation}
where we have set $T_0 = 0$. The time series in Fig.~\ref{fig:method}(b) show the evolution of each term $|   \boldsymbol{\Omega}(t, T_{i-1}) |$ in Eq.~(\ref{eq:new}), with the choice $\theta_T=2$. 

Clearly, the definition in Eq.~(\ref{eq:new}) depends on a novel parameter, the threshold angle $\theta_T$, that sets the series of times $T_i$ at which the threshold is crossed. There are two interesting limiting behaviors. When $\theta_T \to 0$, the threshold is crossed at all timesteps and Eq.~\eqref{eq:new} reduces to the sum of angular increments, and our method then becomes equivalent to the integral method described in Sec.~\ref{sec:angularvelocity}. The opposite limit is when $\theta_T \to \pi$, in which case the angular threshold is never crossed, and Eq.~(\ref{eq:new}) reduces in that case to the Euler vector method described in Sec.~\ref{sec:eulervector}.

These two limits illuminate why the threshold method can be useful. Because it smoothly interpolates between the two conventional methods exposed above, it benefits from their respective advantages while removing their inconsistencies. Let us see why.

First, because the total angular displacement is a discrete sum, Eq.~(\ref{eq:new}), it accumulates angular displacements over discrete periods and this should give rise to an unbounded mean-squared angular displacement, $\Delta \phi^2(t)$, when rotational motion is unconfined. In that case, a Fickian regime is reached at long times that defines a rotational diffusion constant $\Dr$ using Eq.~(\ref{eq:fickian}). The unbounded angular displacement emerging from the threshold method is illustrated in Fig.~\ref{fig:method}(c), to be compared with the bounded Euler vector in  Fig.~\ref{fig:method}(a). 

Second, in the case of a confined random walk the threshold is never crossed if the threshold is chosen larger than the confining angle, $\theta_T > \theta_c$. In that case, the sum in Eq.~(\ref{eq:new}) reduces to its first term, and the Euler vector method is recovered: it works well in that case, recall Fig.~\ref{fig:eulervector}(a). 

\begin{figure}
  \includegraphics[width=\columnwidth]{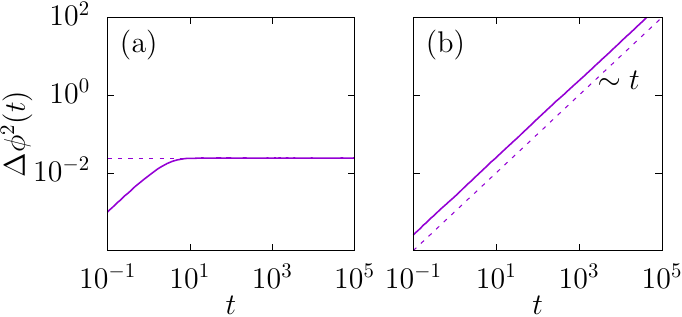}
  \caption{Test of threshold method on (a) a confined and (b) a free angular random walk. The threshold method properly captures both the plateau regime of confined angular motion and the Fickian regime of freely diffusive one.}
  \label{fig:seuiltest}  
\end{figure}

The conclusion is that the threshold method is in principle able to capture the correct physical behaviour in both limits of a confined and a free angular random walk, unlike the conventional methods that failed in either one of these two cases. We now demonstrate numerically that this is the case. In the following (Sec.~\ref{sec:application}), a series of models of increasing complexity is considered to test further the applicability of the threshold method in Eq.~(\ref{eq:new}).

We use Eq.~(\ref{eq:new}) to measure the mean-squared angular displacement $\Delta \phi^2(t)$ in both confined and free angular random walks, see Fig.~\ref{fig:seuiltest}. For the confined random walk in Fig.~\ref{fig:seuiltest}(a) with $\theta_c=0.2$, $\epsilon=0.05$ and $\tau=1$, we set $\theta_T=2.0$. Since the threshold is never crossed, we recover the same (correct) result as in Fig.~\ref{fig:eulervector}(a), without the accumulation of errors at large times observed in Fig.~\ref{fig:testofintegral}(a) for the integral method. 

For the free random walk in Fig.~\ref{fig:seuiltest}(b) with $\tau=1$ and $\epsilon=0.05$, a Fickian behavior is reached at long times, confirming that the total angular displacement in Eq.~(\ref{eq:new}) can grow without bounds. In particular this allows us to extract a rotational diffusion constant $\Dr$. Numerically we find that $\Dr$ depends very weakly on the threshold value $\theta_T$, with the two obvious limits recovered, namely $\Dr = \epsilon^2/ (3 \tau)$ for $\theta_T \to 0$, and $\Dr = 0$ for $\theta_T \geq \pi$. In practice, $\Dr \approx \epsilon^2/ (3 \tau)$ for a broad range of $\theta_T$ values, until it drops abruptly to zero when $\theta_T \to \pi$. (For instance, with $\theta_T = \pi/2$, we obtain  $D_r \approx 0.93 \epsilon^2/ (3 \tau)$, corresponding to a difference of roughly $7\%$ compared to the $\theta_T \to 0$ value). We shall discuss further below how to best adjust $\theta_T$ in more complex situations.  

\section{Continuous time random walk models of rotational motion}

\label{sec:application}

We now introduce a series of models for angular random walks that lead to more complex, slow, or intermittent dynamics. This is done by introducing more complicated choices for the distributions of timescales and angular jumps to explore a broader range of behaviors that are typical of anomalous diffusion in complex and disordered materials. Our goal is to test whether the threshold method introduced in the previous Sec.~\ref{sec:novel} correctly captures and quantifies the physics in all of these cases, that we believe are relevant for the physics of dense molecular fluids.   

\subsection{Intermittent cage escape}

\label{sec:intermittent}

In a first model, we combine fast, confined motion to intermittent escape jumps out of the confined region. This caged dynamics interrupted by rare jumps represents a simplified model that mimics the single molecule intermittent rotational dynamics observed in deeply supercooled molecular liquids~\cite{ediger2000spatially,mandel2022single}. 

The intermittent cage escape model is defined as follows. We start from the confined angular random walk model above. A small random rotation controlled by a parameter $\epsilon$ is performed every $\delta t$ where $\delta t$ is drawn from a Poissonian distribution of average $\tau=1$. For the random rotations, we draw the three components of the rotation vector uniformly in the interval $[-\epsilon,\epsilon]$ and reconstruct the corresponding total rotation matrix using the formalism in Sec.~\ref{sec:methods}. For the confinement, we use the maximum angle $\theta_c$, and reject moves when the accumulated angular rotation becomes larger than $\theta_c$.

\begin{figure}
  \includegraphics[width=\columnwidth]{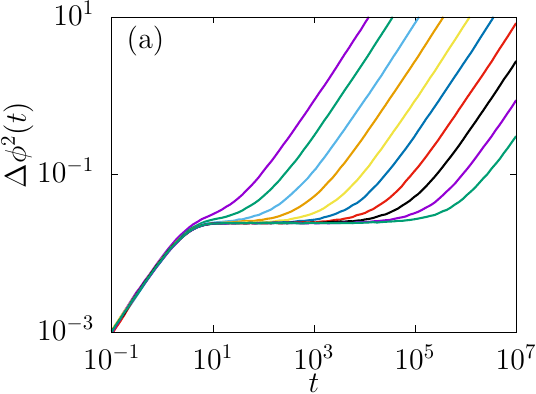}
  \includegraphics[width=\columnwidth]{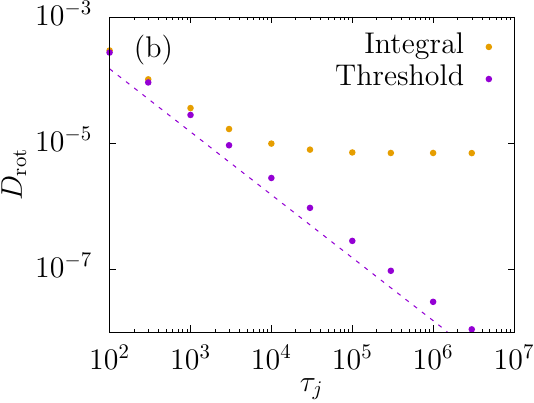}
  \caption{(a) Time dependence of the mean-squared angular displacement for various average escape times $\tau_j$ from $\tau_j=10^2$ to $5 \times 10^6$ using the threshold method.
    (b) The extracted rotational diffusion constant $\Dr$ scales inversely with $\tau_j$ (dashed line). The integral method provides correct results for small $\tau_j$ but leads to erroneous estimates of $\Dr$ at large $\tau_j$, reminiscent of literature results.}
  \label{fig:intermittent}  
\end{figure}

To introduce intermittent escapes out of the cage, we define a second Poissonian distribution of escape times, $P( \delta t) = \tau_j^{-1} \exp(- \delta t /\tau_j)$, with $\tau_j \gg 1$. We draw escape times from $P(\delta t)$ and perform confined random walks in between the jumps as described before. In a jump out of the cage, the system performs a random rotation using again a random rotation vector, with components now drawn from the flat distribution $[-\epsilon_j, \epsilon_j]$, where $\epsilon_j > \epsilon$ controls the typical amplitude of the rotational jumps. We define the position immediately after the jump as the center of the new cage. A confined rotational motion is then restarted from there, until the next cage jump. We show in Fig.~\ref{fig:sketch}(c) a representative trajectory of the system using $\tau_j=4 \times 10^2$ and $\epsilon_j=0.25$. The trajectory qualitatively resembles the behavior observed in deeply supercooled molecular liquids. This model is close, in angular space, to the continuous random walk model proposed in Ref.~\cite{Chaudhuri_2008} in position space.    

To perform quantitative measurements, we implement this intermittent cage escape model, using $\epsilon=0.05$ for the cage amplitude, and $\epsilon_j=0.25$ for the rotational jumps. We are thus left with a single parameter, $\tau_j$, that controls the average time between the cage jumps, that we vary systematically from $\tau_j=10^2$ to $\tau_j=5 \times 10^6$. We use Eq.~(\ref{eq:new}) to reconstruct the total angular displacement vector $\boldsymbol{\phi}(t)$, using a threshold $\theta_T=2.0$. This value is chosen to satisfy both constraints $\theta_c < \theta_T < \pi$ so that confined motion does not lead to an accumulation of small errors to correctly construct an unbounded angular displacement. After averaging over $10^4$ trajectories, we obtain the angular mean-squared displacement $\Delta \phi^2(t)$ shown in Fig.~\ref{fig:intermittent}(a). For each $\tau_j$ we observe the expected two-step dynamics with a rapid approach to a plateau value controlled by the cage size controlled by $\theta_c$, see Eq.~(\ref{eq:thetarandom}), followed by a much slower approach to a Fickian regime, $\Delta \phi^2 (t) \sim t$, that we use to extract a rotational diffusion constant $\Dr$ using Eq.~(\ref{eq:fickian}). When the timescale $\tau_j$ for the cage escape is varied, the angular rotational diffusion becomes slower and the Fickian regime is entered at a longer time. We show in Fig.~\ref{fig:intermittent}(b) that $\Dr$ scales inversely with $\tau_j$, as expected. 

The relevance of the threshold method is illustrated in Fig.~\ref{fig:intermittent}(b) where we report the rotational diffusion constant that would be extracted from the integral method defined in Sec.~\ref{sec:angularvelocity}, and conventionally used in previous simulations of supercooled liquids. In that case the angular mean-squared displacements also become Fickian, but the extracted rotational diffusion constant saturates, for large $\tau_j$, to an incorrect value. This value corresponds to the value measured for the confined random walk in Sec.~\ref{sec:angularvelocity}, and it is dominated by the accumulation of errors discussed in Sec.~\ref{sec:angularvelocity}, and it has no physical meaning. There are several examples of similar observations of a saturation of $\Dr$ in the literature~\cite{chong2009coupling,kawasaki_spurious_2019}. Our analysis demonstrates that they are unphysical due to the mathematical inconsistency of the integral method that assumes the commutation of matrix products. Instead our threshold method properly captures the evolution of $\Dr$ with the control parameters for all explored values.  

\subsection{A model with a slow approach to Fickian regime}

\label{sec:fickian}

Many complex systems are characterized by anomalous diffusion, where the dynamics at intermediate times is neither characterized by a Gaussian distribution of displacements, nor by a Fickian time dependence of mean-squared displacements, and the approach to Fickian/Gaussian dynamics can be very slow~\cite{metzler2000random}. This situation is in particular observed in the translational dynamics of deeply supercooled liquids where Gaussian statistics and Fickian dynamics only arise at very large timescales~\cite{berthier2005length,szamel2006time,berthier2023comment}. While the same physics is expected to occur for rotational motion in molecular supercooled liquids, this has never been studied in detail due to the lack of a correct observable to the long-time rotational motion at very low temperatures. We propose that our threshold method could be fruitfully applied to these situations in the future. 

To demonstrate this point, we construct a continuous time random walk model leading to a transient anomalous diffusive dynamics. We generalize the above model to a non-Poissonian distribution of escape times from the cage. In practice, we use the following normalised distribution 
\begin{equation}
P(\delta t) =
\begin{cases} \displaystyle
\frac{\alpha}{(1+\alpha)\tau_j}
\left( \frac{\tau_j}{ \delta t} \right)^{1+\alpha}, & \delta t \geq \tau_j \\[6pt]
\displaystyle \frac{\alpha}{(1+\alpha)\tau_j}, & 0 \leq \delta t \leq \tau_j ,
\end{cases}
\label{eq:pareto}
\end{equation}
where the expression for times $\delta t<\tau_j$ prevents an unwanted accumulation of very short jump times. The functional form (\ref{eq:pareto}) is interesting because it has a power law decay at large times, controlled by an exponent $\alpha$. This is very close to the Pareto distribution~\cite{mandelbrot1960pareto}. The first and second moment of the distribution $P(\delta t)$ are respectively given by $\langle \delta t \rangle = \tau_j \alpha / [2 (\alpha - 1)]$ (for $\alpha > 1$) and $\langle \delta t^2 \rangle = \tau_j^2 \alpha / [3 (\alpha -2)]$ (for $\alpha > 2$).

When $\alpha$ is very large, the distribution decays very fast for $\delta t > \tau_j$ and the physics is nearly equivalent to the Poisson distribution of jump times used above in Sec.~\ref{sec:intermittent}, with $\langle \delta t \rangle \approx \tau_j/2$ when $\alpha \gg 1$. The physics changes qualitatively when $1 < \alpha < 2$ since the second moment of the distribution and therefore its variance, are infinite. In this regime, the first moment remains finite, with a value that diverges when $\alpha \to 1^+$. Fluctuations are therefore very large, and the dynamics at intermediate times is both strongly non-Gaussian and non-Fickian~\cite{berthier2005length,chaudhuri2007universal}. For times $t \gg \langle \delta t \rangle$ the system should slowly return to Gaussian and Fickian behaviors, and the central limit theorem should eventually apply. We now ask whether such non-trivial behavior can be faithfully characterised when applied to rotational dynamics.

\begin{figure}
  \includegraphics[width=\columnwidth]{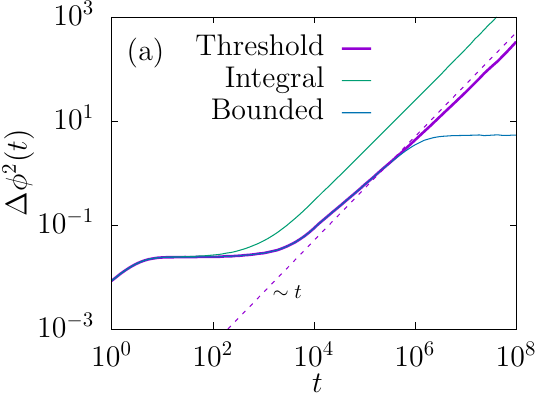}
  \includegraphics[width=\columnwidth]{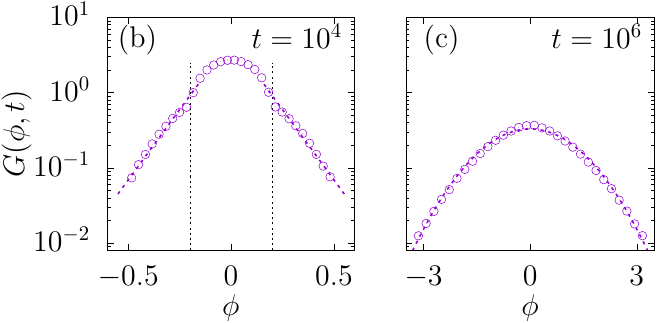}
  \caption{(a) Mean-squared angular displacement for the model described in Sec.~\ref{sec:fickian}, with intermittent cage escapes described by the Pareto distribution (\ref{eq:pareto}) with $\alpha=1.2$. The Fickian regime is reached near $t \approx 10^6$. The rotational diffusion constant is incorrectly estimated by the integral method, and it is not accessible when using the bounded amplitude of the Euler vector. (b, c) Van Hove distributions of angular displacements at two different times. For $t=10^4$ in (b) it is described by the superposition of caged dynamics (vertical dotted lines are $\pm \theta_c$) with exponential tails (dashed lines). For $t=10^6$ in (c) the distribution has nearly converged to its Gaussian limit (dashed line).}   
  \label{fig:CTRW1}  
\end{figure}

We implement this model using $\alpha=1.2$ and $\tau_j=10^4$ for the modified Pareto distribution, keeping for the caged dynamics the Poisson distribution of $\delta t$ with $\tau=1$ and a confinement $\theta_c=0.1$. We run $2 \times 10^3$ trajectories and record the mean-squared angular displacement $\Delta \phi^2(t)$. For selected times, we run a larger number of trajectories (up to $2 \times 10^5$) to measure the van-Hove distributions of the components of the angular displacement vector,
\begin{equation}
  G(\phi,t) = \langle \delta(\phi - \phi_\alpha(t) ) \rangle ,
\end{equation}
averaged over all three directions, $\alpha =x, y, z$.    

The results are shown in Fig.~\ref{fig:CTRW1}. When using the threshold method, we observe that $\Delta \phi^2(t)$ first reaches a plateau, and departs at longer times from the plateau. \rev{A long transient is then observed before Fickian behavior is finally recovered after a time near $t \approx 10^6$. In the long-time limit, a rotational diffusion coefficient $\Dr$ can be measured. We checked that $\Dr \propto \langle \delta t \rangle^{-1} \sim \tau_j^{-1}$ is again satisfied.}

The choice of parameters in Fig.~\ref{fig:CTRW1}(a) is instructive as it illustrates again the challenges for the conventional methods. The integral method correctly captures the plateau regime but the data depart from the plateau too early. This approach provides an incorrect estimate of $\Dr$ and in addition it misses the very slow approach to Fickian dynamics. The approach using the Euler vector also provides the correct dynamics at short times, but it yields a saturation to its long time plateau near $t \approx 10^6$ much before the Fickian regime is entered. From this approach, a determination of $\Dr$ is not possible because the mean-squared angular displacement saturates its bound before the system has entered the Fickian regime, see Fig.~\ref{fig:CTRW1}(a). 

We show in Figs.~\ref{fig:CTRW1}(b, c) the time evolution of the van Hove distributions of angular displacements. For $t=10^4$ in Fig.~\ref{fig:CTRW1}(b), the core of the distribution is controlled by the confined motion within the cage of width $\theta_c$, but the distribution exhibits broad tails that are well described by an exponential decay, $G(\phi,t) \propto \exp ( - \phi/ \lambda)$. The superposition of caged particles and much broader exponential tails reflects the dynamic heterogeneity of the system: at intermediate times most molecules are still confined in their initial cages, but few others have already performed a large number of jumps and are responsible for the exponential tails. At large times, $t=10^6$ in Fig.~\ref{fig:CTRW1}(c), Gaussian behavior is recovered with a width that now grows diffusively, $\propto \sqrt{\Dr t}$.

The situation exposed in Fig.~\ref{fig:CTRW1} is physically very close to findings for translational motion in deeply supercooled liquids, with a slow crossover to Fickian and Gaussian dynamics and the existence of exponential tails in the van Hove distribution of particle displacements. The present analysis of a continuous time random walk model with a modified Pareto distribution for angular dynamics clearly suggests that similar observations could be made for the rotational dynamics of molecular supercooled liquids. The comparison with conventional methods (Euler vector and integral method) also demonstrates that these standard approaches are unable to even attack these questions, whereas the threshold approach that we have developed appears well-suited. This example, together with the choice of model parameters, are physically motivated by observations in deeply supercooled liquids. While the threshold method works well, it relies on a sufficient separation of scales between the cage size $\theta_c$ and the threshold $\theta_T$ that should satisfy $\theta_c < \theta_T < \pi$.

\subsection{Anomalous diffusion with asymptotic sub-diffusive behavior} 

\label{sec:subdiffusion}

\begin{figure}
  \includegraphics[width=\columnwidth]{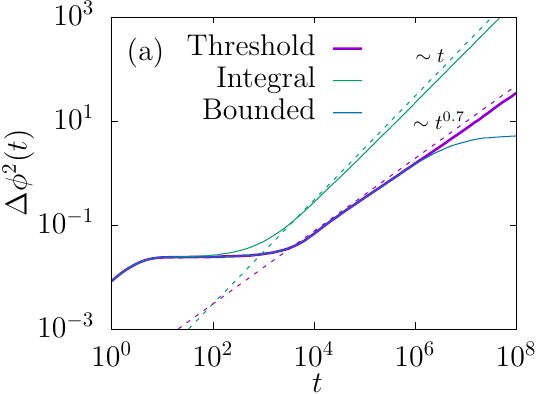}
  \includegraphics[width=\columnwidth]{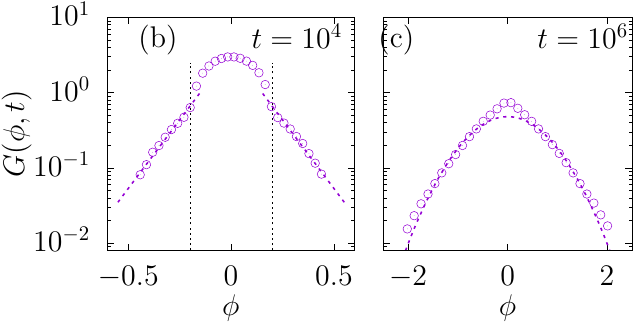}
  \caption{(a) Mean-squared angular displacement for the model described in Sec.~\ref{sec:subdiffusion}, with intermittent cage escapes described by the Pareto distribution (\ref{eq:pareto}) with $\alpha=0.7$. A sub-diffusive regime is reached at large times with $\Delta \phi^2 \sim t^{\alpha}$. The integral method predicts an incorrect Fickian regime, while a determination of the sub-diffusive exponent is impossible using the Euler vector. (b, c) Van Hove distributions of angular displacements at two different times. For $t=10^4$ in (b) it is described by the superposition of caged dynamics (vertical dotted lines are $\pm \theta_c$) with exponential tails (dashed lines). For $t=10^6$ the distribution has nearly converged to its non-Gaussian asymptotic limit (the dashed line is a Gaussian distribution, for comparison).}
  \label{fig:CTRW2}  
\end{figure}

We now study the case where the exponent $\alpha$ in the modified Pareto distribution becomes smaller than $1$. In that case, the first moment of the distribution is divergent, and the random walk process never converges to a Fickian regime with a Gaussian distribution of displacements, and the central limit theorem never applies. Instead the mean-squared angular displacement is expected to display sub-diffusive behavior at large times with an exponent controlled by $\alpha$, associated to universal non-Gaussian statistics of angular displacements~\cite{metzler2000random}.

This case is studied in Fig.~\ref{fig:CTRW2} where we show the mean-squared angular displacement averaged over $2 \times 10^3$ trajectories, with $\alpha=0.7$ and $\tau_j = 10^4$. The parameters for the caged dynamics at short times are as before, see Sec.~\ref{sec:fickian}. The overall behavior of $\Delta \phi^2(t)$ is as expected, with a fast approach to the plateau followed by a long transient towards an asymptotic sub-diffusive behavior $\Delta \phi^2 \sim t^\alpha$. The integral method is again totally inaccurate at large times. It continues to predict an incorrect Fickian behavior at large times, and qualitatively misses the asymptotic sub-diffusive regime. The Euler vector method yields bounded dynamics towards a long-time plateau that prevents the determination of the sub-diffusion exponent. The van Hove distributions in Figs.~\ref{fig:CTRW2}(b, c) also display an interesting time evolution, from the superposition of caged dynamics with exponential tails at $t=10^4$, to an asymptotic non-Gaussian distribution in the sub-diffusive regime for $t=10^6$. 

The slow sub-diffusive behavior shown in Fig.~\ref{fig:CTRW2} mimics the aging dynamics of supercooled liquids after a sudden quench to the glass phase, where the average relaxation time is so large that Fickian dynamics is never observed. We believe that the observables introduced here would be very valuable to analyse the rotational dynamics of aging molecular fluids. Once again, none of the existing methods would be able to correctly capture the statistical features of these non-trivial rotational dynamics. The separation of scales between $\theta_c$, $\theta_T$ and $\pi$ is again needed for the threshold method to work efficiently.  

\section{Conclusion and outlook}

\label{sec:conclusion}

We have discussed state-of-the-art methods to quantify the long-time rotational dynamics of molecular systems, with a particular interest for the definition of a rotational diffusion constant $\Dr$ describing an asymptotic Fickian regime. While we considered very simple models, our motivation to perform such analysis stems from the physics of dense molecular fluids in which molecules perform slow, heterogeneous and intermittent rotational dynamics~\cite{ediger2000spatially}. The connection (or decoupling) between rotational and translational dynamics in such systems is a broad area of research, also associated to the possible breakdown of the Stokes-Einstein-Debye relation.  

We have shown that the two main families of existing methods to describe the long-time rotational dynamics of molecular systems encounter serious limitations and may even provide qualitatively incorrect or limited results when applied to systems with slow and intermittent dynamics.

In short, the method using the total rotation vector correctly locates the orientation of the molecule at any time, but is unable to provide an unbounded angular displacement that would display an asymptotic Fickian regime with Gaussian distributions of angular displacements. While this may be fine for simple systems described by Brownian angular diffusion, this raises serious obstacles to analyse the statistics of rotational motion in systems with more complex rotational dynamics, long transient behaviors, or non-Fickian asymptotic limits. 

The alternative method that integrates the angular velocity to provide an angular displacement incorrectly locates the orientation of molecules at large times, because the total angular displacement is not equal to the total rotation matrix. While it has been noticed that this method provides rotational diffusion constant that do not slow down as much as was physically expected~\cite{chong2009coupling,kawasaki_spurious_2019}, we showed that the rotational diffusion constant extracted using this method in fact does not reflect the real rotational motion of the molecules. In particular, $\Dr$ remains finite even in glasses where molecular rotations are completely arrested. Deep down, the error made in this approach stems from the basic mathematical fact that rotation matrices do not commute. This is of course a well-known fact, but its consequences on the determination of $\Dr$ using the integral method had not been appreciated. 

\rev{We did not discuss methods based on projections of the orientation of the molecules on a specific set of axis, such as the vertical axis or a horizontal plane. In the same vein, the orientation of a molecule can be partially tracked by following its longitude or latitude on the sphere, or alternatively using Euler angles. By reducing the three-dimensional information, it would seem these methods solve the problem of non-commuting rotation matrices, so that accumulating small displacements becomes mathematically well-defined. Another problem however prevents the use of all these approaches. Take for instance the $xy$ projection of a three-dimensional polarisation vector that defines a rotation angle in the plane. When using this angular observable, a problem arises when the polarisation vector is nearly aligned with the $z$ axis, as small rotations in this position may lead to multiple $2\pi$ turns in the angle defined on the plane. Lack of three-dimensional information in fact always leads to similar problems, and the angular variables then do not lead to the correct Fickian dynamics at long times, and systematically overestimates the orientational dynamics.}        

To properly characterise the asymptotic angular dynamics of molecules in deeply supercooled liquids, we have introduced a novel, empirical method that provides correct results in all studied cases: free and confined diffusion, anomalous dynamics with slow approach to Fickian regime and anomalous sub-diffusion. We showed that our threshold method combines the advantages of both bounded and integral methods, while removing their shortcomings. This method provides accurate results in all of these cases. It allows us to access the whole distribution of angular displacements, resolved in all three directions. We could thus construct van Hove angular distributions and a mean-squared angular displacement, in complete analogy with translational dynamics. \rev{For the models studied here, where parameters were set to reflect semi-quantitatively the physics of supercooled molecular fluids, we found in particular that the specific choice of the threshold parameter $\theta_T$ has little quantitative influence on the extracted value of $\Dr$. This remains to be confirmed in the analysis of realistic molecular models.} 

While this approach works for all the theoretical models proposed here, an obvious limitation would arise when the separation of scales between short-time dynamics and the threshold parameter ceases to exist, or when the distribution of instantaneous angular jumps becomes too large, as for a model for angular L\'evy walks for instance, although physical relevance for molecular fluids is not obvious. \rev{It would be interesting to construct synthetic models or analyse physical situations where this separation of scales is lost. We suspect that alternative methods would be needed for these situations.}  

Despite these limitations, we believe that our threshold method will be very useful to study molecular fluids approaching the glass transition. In a forthcoming publication, we will apply this new method to get novel insight into the temperature evolution of the rotational diffusion constant, the distribution of angular displacements, and its possible coupling or decoupling to translational motion, and we shall revisit the validity of the Debye-Stokes-Einstein relation in deeply supercooled molecular fluids. \rev{Our results suggest a rotational diffusion coefficient $\Dr(T)$ can be determined in these systems and will not be affected by the limits of the integral method. As a result $\Dr(T)$ should represent a novel transport coefficient with an interesting temperature evolution, which may or may not couple to other rotational or translational timescales.}  

\section*{Data Availability}

The data that support the findings of this study are openly available on Zenodo~\cite{simon_2026_20026328}.

\acknowledgments  

We thank K. Bendriss, P.-M. D\'ejardin, D. Frenkel, R. Jack, J. Kurchan, V. Lecomte, and G. Tarjus for many interesting discussions. L.B. acknowledges the support of the French Agence Nationale de la Recherche (ANR), under grants ANR-20-CE30-0031 (project THEMA) and ANR-24-CE30-0442 (project GLASSGO).

\bibliography{sample}

\newpage

\appendix

\section{Fokker Planck equation for random walks in $\mathrm{SO}(3)$}

\label{app:FP}

Equation (\ref{eq:rottime}) describes the successive composition of infinitesimal rotations. In the axis-angle representation, and in the limit $\epsilon \rightarrow 0$, the corresponding stochastic process can be seen as the discretization of the following stochastic differential equations, interpreted  according to the It\^o rule:
\begin{equation}
\dot u = - \frac{3 \sigma^2}{8}u -\frac{1}{2} \sqrt{1-u^2} \mathbf{n}\cdot\mathbf{w}
\label{Ito1}
\end{equation}
\begin{equation}
 \mathbf  {\dot n} = -\frac{\sigma^2}{4(1-u^2)} \mathbf{n}+  \frac{1}{2\sqrt{1-u^2}} \left(\mathbf{w}-(\mathbf{n}\cdot\mathbf{w})\mathbf{n}\right) + \frac{1}{2}  \mathbf{w}\times \mathbf{n}
 \label{Ito2}
\end{equation}
with $u= \cos (\theta/2)$,  where $\mathbf{w}= (w_1,w_2,w_3)$ and $w_i$ a white noise with variance $\sigma^2$.

As a result, the distribution $P(u,\mathbf{n})$ evolves according to the following Fokker-Planck equation:
\begin{equation}
\frac{\partial P}{\partial t}=\sigma^2 \frac{\partial }{\partial u}\left(  \frac{3u}{8}P\right)+\sigma^2  \frac{\partial^2 }{\partial u^2} \left( \frac{1-u^2}{8}P\right)+ D_{\mathrm{eff}}(u)\Delta_{\mathbf{n}}P
\end{equation}
where $\Delta_{\mathbf{n}}$ is the Beltrami Laplacian acting on the unit sphere, and  $D_{\mathrm{eff}}(u)= \frac{\sigma^2}{4 ({1-u^2})} $. The stationary solution for $P(u,\mathbf{n})$ is independent of $\mathbf{n}$ and proportional to  $\sqrt{1-u^2}$. While this uniform measure is well known in the mathematical literature, we are not aware of a previous derivation using the Fokker-Planck equation to obtain it.

Note that although the vector $\mathbf{n}$ is uniformly distributed in the stationary state, the corresponding stochastic process is not a standard diffusion on the unit sphere, as the effective diffusion constant depends on the angle.

In order to prove the above, one may start from the Rodrigues formula for the composition of two rotations $\alpha \mathbf{l}$ and $\beta \mathbf{m}$, which is a rotation $\gamma  \mathbf{n}$. One has~\cite{rodrigues1840lois,Altmann1989}: 
\begin{equation}
\cos  \frac{\gamma}{2}=\cos \frac{\alpha}{2} \cos \frac{\beta}{2} - \sin \frac{\alpha}{2} \sin \frac{\beta}{ 2 } \ \   \mathbf{l}\cdot \mathbf{m} ,
\end{equation}
and
\begin{equation}
\sin \frac{\gamma}{ 2}\  \mathbf{n} = \sin \frac{\alpha}{ 2}  \cos \frac{ \beta }{ 2 }\ \mathbf{l} + \cos \frac{\alpha}{ 2} \sin \frac{ \beta }{ 2 }\ \mathbf{m} + \sin \frac{\alpha}{ 2}  \sin \frac{ \beta}{ 2 } \ \ \mathbf{l}\times \mathbf{m} .
\end{equation}

We now specialize to the case where the second rotation is  a small increment, $\beta \mathbf{m} = (w_1,w_2,w_3)$,  where $w_i$ is a white noise with variance $\sigma^2$, and consider $\alpha \mathbf{l}$ and $\gamma  \mathbf{n}$ as two successive steps in the random walk described in the main text, Eq.~(\ref{eq:rottime}), $\alpha \mathbf{l}=\theta_k\mathbf{n}_k  $, $\gamma  \mathbf{l}=\theta_{k+1} \mathbf{n}_{k+1}$. Using $u_k=\cos({\theta_k/2})$  one has easily:
\begin{equation}
u_{k+1}-u_k = - u_k \frac{\beta^2}{ 8} - \frac{1 }{ 2}\sqrt{1-u_k^2}\  \beta  \ \mathbf{m}\cdot \mathbf{n}_k  ,
\end{equation}
which, using that ${ \overline {\beta^2} = 3\sigma^2}$  is the It\^o discretization of Eq.~(\ref{Ito1}). For the evolution of the rotation vector, one obtains
\begin{eqnarray}
\displaystyle \mathbf{n}_{k+1} -\mathbf{n}_k  = & - \frac{\beta^2 \vert \mathbf{m}_{\perp}\vert^2}{8(1-u_k^2)} \mathbf{n}_k  + \frac{u_k}{\sqrt{1-u_k^2}} \frac{\beta}{ 2} \left( \mathbf{m} - (\mathbf{n_k} \cdot \mathbf{m})  \mathbf{n_k}\right) \nonumber \\ &
+ \frac{\beta}{ 2} \mathbf{n}_k\times  \mathbf{m} .
\end{eqnarray}
The first drift term ensures that $\vert \mathbf{n}_{k+1} \vert=1$, with $ \mathbf{m}_{\perp}$ the projection of $\mathbf{m}$ perpendicular to $\mathbf{n}$. Replacing $\beta^2$ by $3\sigma^2$ and  $\vert \mathbf{m}_{\perp}\vert^2$ by $2/3$, this  corresponds to the It\^o discretization of Eq.~(\ref{Ito2}).

\end{document}